\documentclass[preprint,12pt]{elsarticle}
  \newcommand{\bsll}{$b \rightarrow s l^+ l^-$}

\usepackage{amssymb}

\journal{Physics Letters B}

\begin{document}

\begin{frontmatter}



\title{LHC di-lepton searches for $Z^\prime$ bosons which explain measurements
  of \bsll\ transitions} 

\author[1]{Ben Allanach}
\ead{ben.allanach.work@gmail.com}
\affiliation[1]{organization={DAMTP, University of Cambridge},
  addressline={Centre for Mathematical Sciences, Wilberforce Road},
  postcode={CB3 0WA},
  city={Cambridge},
  country={United Kingdom}}


\begin{abstract}
Several current measurements of \bsll\ processes are in tension with Standard
Model predictions, whereas others (e.g.\ $R_K$ and $R_{K^\ast}$) are in
reasonable agreement with them. We examine some recent $Z^\prime$ models that
fit the data as a whole appreciably better than does the Standard Model,
confronting the models with ATLAS resonance searches in the $e^+e^-$ and
$\mu^+\mu^-$ channels. Both channels constrain the models; we find that the
searches rule out less than one fifth of the allowed range of
$M_{Z^\prime}$. We estimate that the HL-LHC can roughly double the current
sensitivity in $M_{Z^\prime}$, but covering the whole parameter space would require a more powerful machine, such as a 100 TeV $pp$ collider.
\end{abstract}




\end{frontmatter}



Some experimental measurements of $B-$meson decays
involving the \bsll\ transition
are in tension with
Standard Model (SM) predictions~\cite{Gubernari:2022hxn}. Focusing on
the most discrepant measurements, 
the branching ratio $BR(B_s
\rightarrow \phi \mu^+\mu^-)$, for example,  is measured~\cite{LHCb:2021zwz}
to be 4$\sigma$ 
below SM predictions. Some angular distributions in $B \rightarrow K^\ast
\mu^+\mu^-$ decays are also around 4$\sigma$~\cite{LHCb:2015svh,ATLAS:2018gqc,CMS:2017rzx},
as is $BR(B \rightarrow K \mu^+ \mu^-)$~\cite{Parrott:2022zte}.
On the other hand, many other measurements involving the $b\rightarrow s
l^+l^-$ transition are compatible with SM predictions, notably 
$R_K$ and $R_{K^\ast}$, which are ratios of branching ratios
$$
R_M = \frac{BR(B \rightarrow M \mu^+\mu^-)}{BR(B \rightarrow M e^+ e^-)},
$$
for $M \in \{ K, K^\ast \}$~\cite{LHCb:2022qnv}. 
Although the aforementioned more discrepant observables do suffer larger
theoretical 
errors, several estimates (e.g.\ Refs.~\cite{Gubernari:2022hxn,Parrott:2022zte}) imply that
theoretical uncertainties are insufficient to explain the tensions whilst respecting other measurements. 

In Ref.~\cite{Allanach:2023uxz}, the SM was extended in order to ameliorate
the tensions. An 
additional 
spontaneously broken $U(1)_X$ gauge group with family-dependent charges is
posited to be in a direct product with the SM gauge group. The resulting
additional TeV-scale vector boson, dubbed a $Z^\prime$ boson, mediates \bsll\ 
transitions, altering predictions for their observables. The
$U(1)_X$ charge 
assignment of the SM fermionic fields (including three right-handed neutrino fields) is 
\begin{equation}
  X = 3B_3 - (X_e L_e + X_\mu L_\mu + [3-X_e-X_\mu] L_\tau), \label{charges}
\end{equation}
where $B_3$ is third-family baryon number, $L_e,L_\mu$ and $L_\tau$ are
first, second and third family lepton number, respectively and $X_e, X_\mu \in
\mathbb{Q}$.
These simple bottom-up models~\cite{Allanach:2023uxz} explain why $|V_{td}|$, $|V_{ts}|$,
$|V_{ub}|$ and $|V_{cb}|$ are 
all small compared to unity and why charged lepton flavour violation is
suppressed. 
The assignment (\ref{charges})
is free of perturbative gauge anomalies, a necessary condition for a consistent
theory. 
Previously, when it was thought that $R_K$ and 
$R_{K^\ast}$ differed more significantly from their SM predictions than 
updated analyses suggest, a $Z^\prime$ model with 
$X_\mu=3$, $X_e=0$ was
proposed~\cite{Bonilla:2017lsq,Alonso:2017uky,Allanach:2020kss}. The case
$X_e=X_\mu=X_\tau=1$ was examined~\cite{Greljo:2022jac} more recently. 

\begin{figure}
  \begin{center}
    \unitlength=\textwidth
    \begin{picture}(0.99, 0.3)
      \put(-0.07,-0.08){\includegraphics[width=0.7 \textwidth]{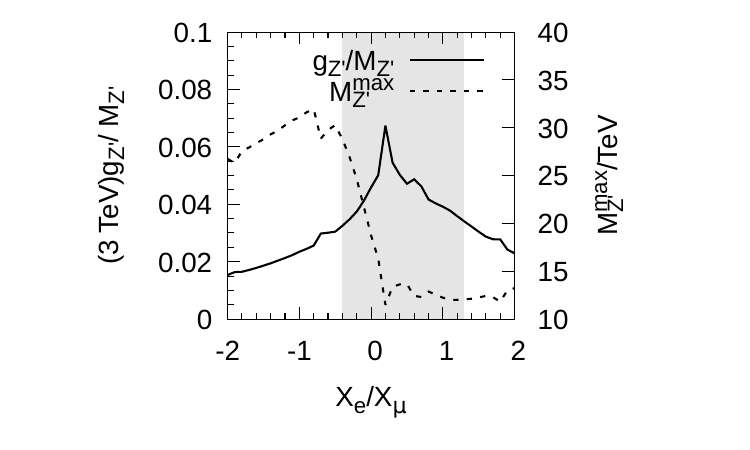}}
      \put(0.40,-0.08){\includegraphics[width=0.7 \textwidth]{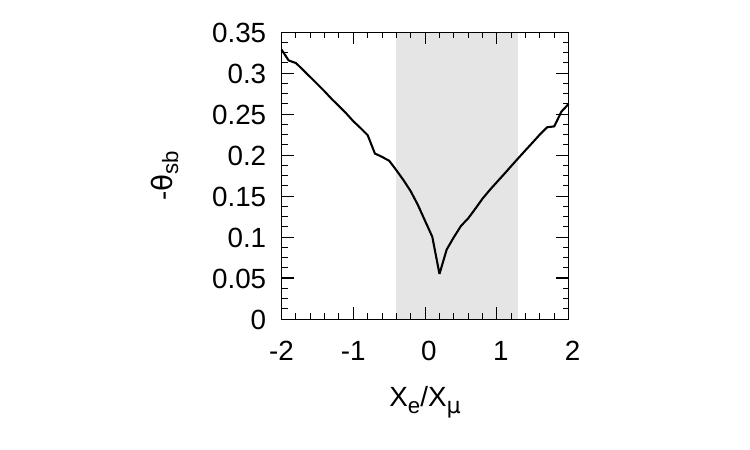}}
      \put(0.0,0.3){(a)}
      \put(0.5,0.3){(b)}
    \end{picture}
  \end{center}
  \caption{\label{fig:params} Best-fit model parameters along the model line
    parameterised by $X_e/X_\mu$ from the fit to flavour and LEP data in
    Ref.~\cite{Allanach:2023uxz}, for $X_\mu=10$ and $M_{Z^\prime}=3$ TeV.  
    The 95$\%$ CL region favoured by the fit is
    shaded. $M_{Z^\prime}^\textrm{\footnotesize max}$ is an estimate of the maximum value of
    $M_{Z^\prime}$ coming from perturbativity, explained just under
    (\ref{width}).} 
\end{figure}

In a global fit to 247 measurements of processes involving the $b \rightarrow
s$ transition and 148 
LEP-2 $e^+ e^-$ to di-lepton data, Ref.~\cite{Allanach:2023uxz} showed that
the $U(1)_X$ models could provide a better fit
than the SM by some 16 units of $\chi^2$ and a reasonable goodness-of-fit with
a $p-$value around $0.2-0.3$. The fit is sensitive 
to
the ratio of the $Z^\prime$ gauge coupling to the $Z^\prime$ mass
$x:=g_{Z^\prime}/M_{Z^\prime}$, $X_e$, $X_\mu$ and
$\theta_{sb}$, a mixing angle between $s_L$
and $b_L$. For each value of $X_e$ and $X_\mu$,
a fit was performed to $x$ and $\theta_{sb}$.
It was found, assuming the hypothesis of the line of models parameterised
by $X_e/X_\mu$, that $-0.4 \leq X_e/X_\mu\leq 1.3$ to 95$\%$ confidence level (CL). We show the
best-fit values of parameters from the fit as a function of
$X_e/X_\mu$ in Fig.~\ref{fig:params}, for $X_\mu=10$: these we shall use
throughout the rest of this letter. 
The parameters are shown for $M_{Z^\prime}=3$ TeV and
$X_\mu=10$, although it is indicated on the left-hand ordinate of Fig.~\ref{fig:params}a
how (to a good approximation) the best-fit gauge coupling scales\footnote{This
scaling comes from the fact that all dimension-6 SMEFT 
operators induced by integrating out the $Z^\prime$ boson are proportional to
$g_{Z^\prime}^2/M_{Z^\prime}^2$.} as a function of $M_{Z^\prime}$. 

Here for the first time, we re-cast direct $Z^\prime$ search results
in this line of   
models. Such searches at the LHC have so far reported no significant excess
and so we expect there to be concomitant constraints impinging on the
parameter space of each model along the model line. We shall also provide
rough sensitivity estimates for the high luminosity (HL-LHC) run for such
searches. 

In the models investigated here, 
the dominant production mode\footnote{The next largest production mode
production cross-section $(s \bar b+\bar b s) \rightarrow Z^\prime$
is smaller than that of the $b \bar b$
production by a factor which depends on parameters, but is here always more
than 20.} of the $Z^\prime$ boson in $pp$ 
collisions at the 
LHC is via $b \bar b \rightarrow Z^\prime$
and thus the $Z^\prime$ production
cross-section $\sigma$ is suppressed by the small bottom parton
distribution function squared, but otherwise
\begin{equation}
  \sigma \propto g_{Z^\prime}^2 \cos^4 \theta_{sb} = g_{Z^\prime}^2 (1 - 2
  \theta_{sb}^2 + {\mathcal O}(\theta_{sb}^4) ). \label{prod}
\end{equation}
The production cross-section is therefore approximately independent of
$\theta_{sb}$ if its value is significantly smaller than unity (this is the case
for our model line) and 
indeed we shall neglect the dependence\footnote{We have checked that setting
$\theta_{sb}=0$ throughout instead makes scant difference to our results.} of $\theta_{sb}$ upon $M_{Z^\prime}$.

A $Z^\prime$ decays into a fermion and an
anti-fermion, to a good approximation.
The partial width of a $Z^\prime$ boson decaying into a Weyl fermion $f_i$ and  
a Weyl anti-fermion $\bar f_j$ is
\begin{equation}
  \Gamma_{ij} = \frac{C}{24 \pi} |g_{ij}|^2 M_{Z^\prime},
\end{equation}
where $g_{ij}$ is the coupling of the $Z^\prime$ boson to $f_i \bar f_j$ and
$C$ is the number of colour degrees of freedom of the fermions (here, 3 or
1). In the limit that $M_{Z^\prime} \gg 2m_t$, we may approximate all fermions
as being massless, save for the right-handed neutrinos, which we assume are
more massive than $M_{Z^\prime}/2$ and so are kinematically impossible for an on-shell
$Z^\prime$ to decay into. We also neglect the small effects in decay widths coming from quark mixing. 
We show the various partial widths as functions of $X_e$ and $X_\mu$ in
Table~\ref{tab:pws}. 
\begin{table}
  \begin{center}
    \begin{tabular}{|c|cccccccc|} \hline
  $ij$    & $b \bar b$ & $t \bar t$ & $e^+e^-$ & $\mu^+\mu^-$ & $\tau^+ \tau^-$  &
      $\nu_e \bar \nu_e$ & $\nu_\mu \bar \nu_\mu$ & $\nu_\tau \bar \nu_\tau$ 
      \\ \hline
      $\frac{\Gamma_{ij}}{M_{Z^\prime} g_{Z^\prime}^2 \pi}$ & $\frac{1}{4}$ &
      $\frac{1}{4}$ & $\frac{X_e^2}{12}$ & $\frac{X_{\mu}^2}{12}$ &
      $\frac{(3-X_e-X_\mu)^2}{12}$ & $\frac{X_e^2}{24}$ & $\frac{X_{\mu}^2}{24}$ &
      $\frac{(3-X_e-X_\mu)^2}{24}$\\  \hline
    \end{tabular}
  \end{center}
  \caption{\label{tab:pws} Partial widths of the $Z^\prime$ into various
    fermion/anti-fermion pairs in the massless unmixed
    fermion approximation.}
\end{table}
Summing all of these, we obtain a total $Z^\prime$ width of
\begin{equation}
  \Gamma = \frac{4 + X_e^2 + X_\mu^2 + (3-X_e-X_\mu)^2}{8 \pi} g_{Z^\prime}^2
  M_{Z^\prime}. \label{width}
\end{equation}
Perturbativity requires that the $Z^\prime$ is not too broad a resonance
(here, we use
$\Gamma/M_{Z^\prime} < \frac{1}{2}$),
yielding an upper limit upon $g_{Z^\prime}$ from (\ref{width}). Substituting
this value 
into $g_{Z^\prime}/M_{Z^\prime}$ from Fig.~\ref{fig:params}a then yields a maximum
value (which we call $M_{Z^\prime}^\textrm{\footnotesize max}$) of $M_{Z^\prime}$ that can still provide a best-fit to the flavour and
LEP data while maintaining perturbativity,
as shown in Fig.~\ref{fig:params}a. 

\begin{figure}
  \begin{center}
    \unitlength=\textwidth
    \begin{picture}(0.99, 0.37)
      \put(-0.15,-0.1){\includegraphics[width=0.85 \textwidth]{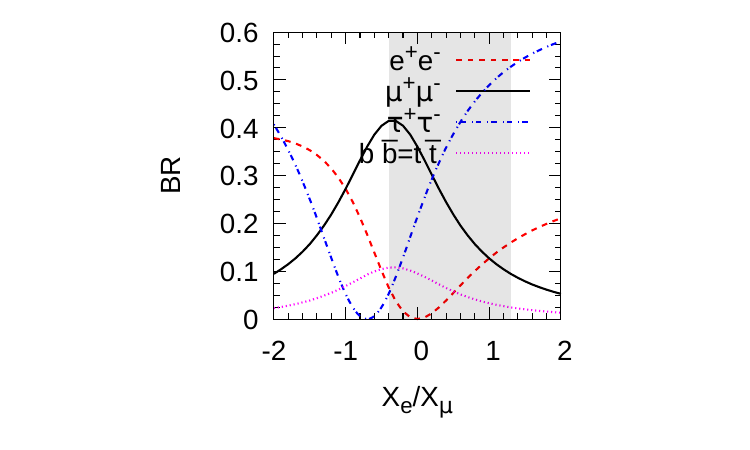}}
      \put(0.35,-0.1){\includegraphics[width=0.85 \textwidth]{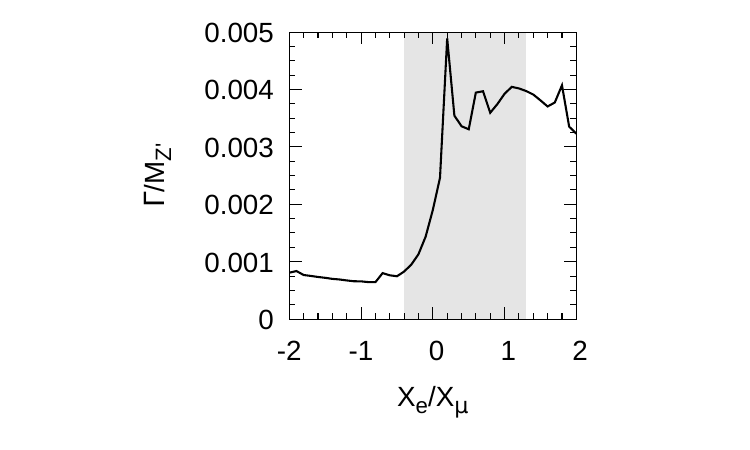}}
      \put(0.0,0.37){(a)}
      \put(0.5,0.37){(b)}
    \end{picture}
  \end{center}
  \caption{\label{fig:brs} $Z^\prime$ decays for $M_{Z^\prime} \gg 2m_t$ and
    $X_\mu=10$ resulting from the fit to flavour and LEP-2 data in
    Ref.~\cite{Allanach:2023uxz}: (a) predicted
    $Z^\prime$ branching ratios into 
    observable final states. (b) width over mass for $M_{Z^\prime}=3$ TeV. 
    The 95$\%$ CL region favoured by the fit  is shaded.}
  \end{figure}

We display the values of some branching ratios into visible final-state
particles in Fig.~\ref{fig:brs}a for the case $X_\mu=10$. 
  We see from the figure that $\mu^+\mu^-$ and/or $e^+e^-$ final states have an 
appreciable branching ratio. 
The LHC experiments perform searches in these
by looking for a signal peak on top of a smooth background distribution in the di-lepton 
mass. Experimentally, di-leptons enjoy much higher efficiencies and/or lower backgrounds than 
$t \bar t$, $b \bar b$ or $\tau^+ \tau^-$ 
and so
we expect the $\mu^+ \mu^-$ and $e^+e^-$ channels to provide the best
sensitivity, 
since there is no huge branching ratio factor favouring the other
ones. From now on, we take `di-lepton' states to mean $\mu^+ \mu^-$ and
$e^+e^-$. 

In Ref.~\cite{ATLAS:2019erb}, ATLAS reported a search for heavy resonances in
di-lepton 
invariant mass distributions for 139 fb$^{-1}$ of 13 TeV $pp$ collisions at
the LHC\@. Having observed no significant excess, 
ATLAS provides a 95$\%$ CL upper bound  on production cross section
multiplied 
by branching 
ratio of a new physics state decaying into the relevant di-lepton final state
depending upon $z$, defined to be the ratio of the total width of the new
physics state divided by its mass. We call this bound $s(z, M_{Z^\prime})$. 
ATLAS provides bounds both for a narrow resonance 
($s(0, M_{Z^\prime})$) and for
a resonance whose width is a tenth of its mass ($s(0.1, M_{Z^\prime})$), as
well as some particular discrete values of $z \in (0, 0.1)$.
In 
Ref.~\cite{Allanach:2019mfl}, it was shown that for the di-muon final state, the function
\begin{equation}
  s(z, M_{Z^\prime}) = s(0, M_{Z^\prime})  \left( \frac{s(0.1,
    M_{Z^\prime})}{s(0, M_{Z^\prime})}
    \right)^\frac{z}{0.1} \label{interp}
  \end{equation}
provides a reasonable interpolation for the bounds at other values of $z$
between 0 and 0.1. We shall use (\ref{interp}) for the di-muon channel and also extend its
use to the di-electron channel.
\begin{figure}
  \begin{center}
    \unitlength=\textwidth
    \begin{picture}(0.99, 0.3)
    \put(0,0){\includegraphics[width=0.49\textwidth]{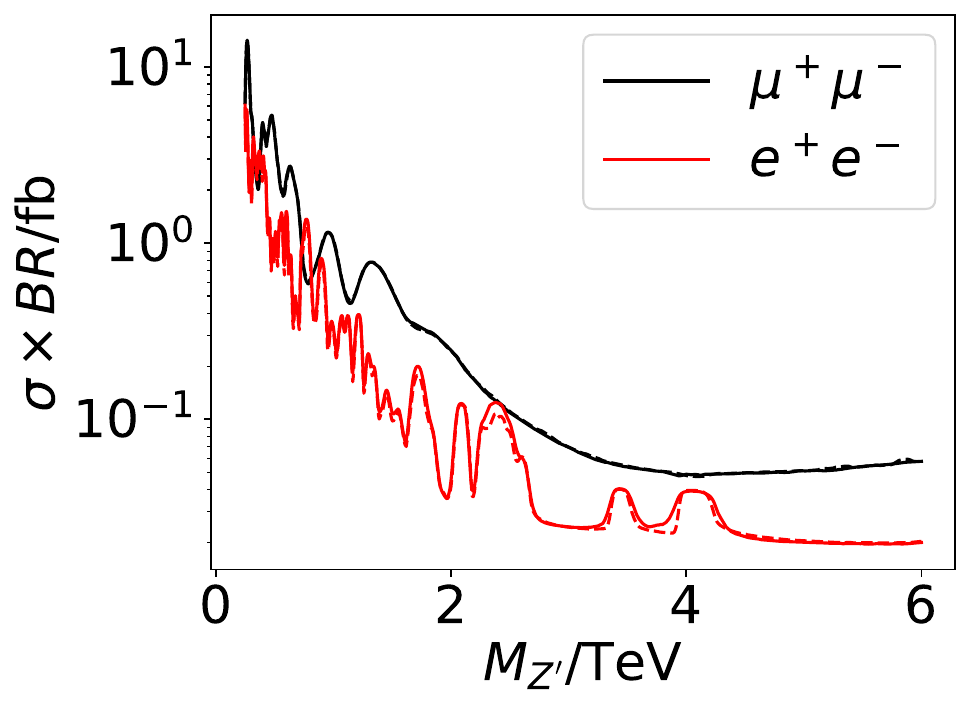}}
    \put(0.5,0){\includegraphics[width=0.49\textwidth]{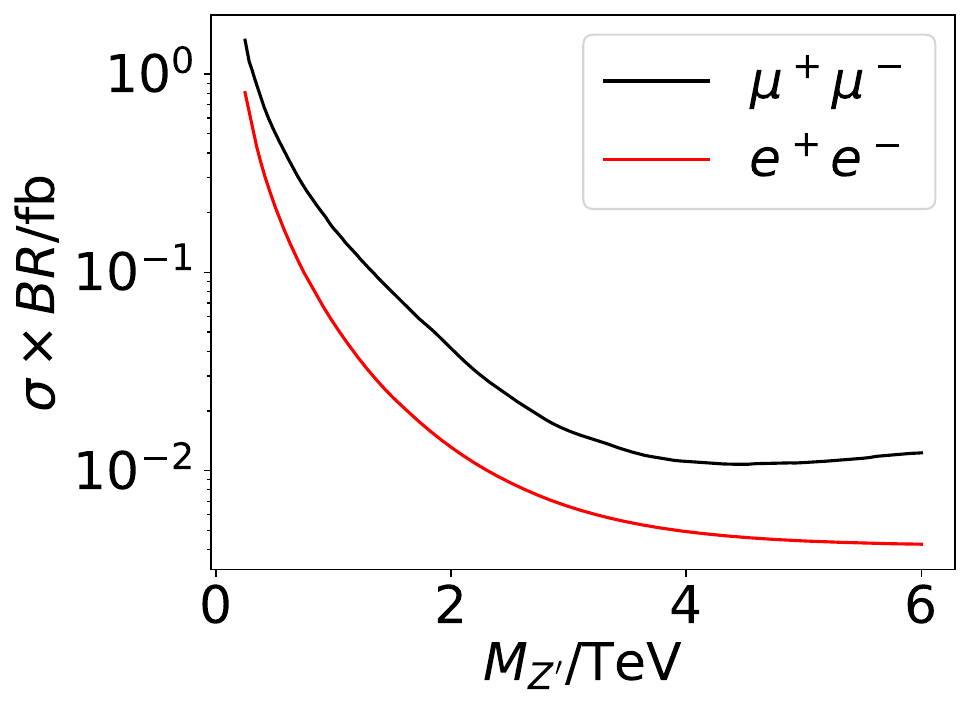}}
      \put(0.0,0.33){(a)}
      \put(0.5,0.33){(b)}
    \end{picture}
  \end{center}
  \caption{\label{fig:val} ATLAS
    bounds from resonant di-lepton searches.
    (a) Validation of our interpolation: the 
    solid lines show the 
    95$\%$ CL upper bound for $\Gamma/M_{Z^\prime}=0.012$ observed by
    ATLAS~\cite{ATLAS:2019erb}  
    whereas our interpolations from (\ref{interp}) are shown by dashed
    curves. Where a dashed curve 
    is not visible, it coincides with a solid     one, indicating an excellent
    approximation. (b) Estimated HL-LHC sensitivity for a narrow
    $Z^\prime$.}  
\end{figure}
As Fig.~\ref{fig:brs}b shows, for the
parameter set taken here ($X_\mu=10$), the
$Z^\prime$ boson
is narrow for $M_{Z^\prime}=3$ TeV. In Fig.~\ref{fig:val}a, we validate
(\ref{interp}) with data  
provided by the ATLAS collaboration at $z=0.012$. The proximity of the dashed
and solid curves in the figure demonstrates that the 
interpolation (\ref{interp}) also works for the $e^+e^-$ channel. We
compute the production cross-section times 
branching ratio at tree-level order with {\tt
  MadGraph3.5.3}~\cite{Alwall:2014hca}, having encoded the model in {\tt UFO}
format~\cite{Degrande:2011ua}.  

We shall provide a rough and conservative estimate of the sensitivity of the
HL-LHC di-lepton searches. We assume an integrated  luminosity of 3000
fb$^{-1}$. It is convenient for us to estimate the sensitivity assuming the same centre-of-mass energy
as Run I (13 TeV); although the energy and future running schedule is unknown,
it is likely to be at a slightly higher energy (indeed, the latest LHC run has
been recording $pp$
collisions at 13.6 TeV). We may expect higher centre-of-mass energies to have
higher sensitivities to TeV-scale heavy resonant states, since the production
cross-sections increase but the backgrounds are low.
We may therefore expect the HL-LHC sensitivity to be similar but
slightly higher than a prediction based on a slightly lower energy. Given the
large uncertainty in the final 
integrated luminosity, we deem our rough estimate sufficient.
When backgrounds are low as is the case for larger values of $M_{Z^\prime}$,
the sensitivity scales as the square root of the expected number of signal
events $\sqrt{S} \propto\sqrt{\mathcal L}$. 
Even when the expected number of background events $B$ is \emph{not} low, 
the sensitivity of a search roughly scales as $S/\sqrt{B}$ and since both $S$
and $B$ are proportional to the integrated 
luminosity collected ${\mathcal L}$, we still expect the sensitivity to scale
proportional to $\sqrt{\mathcal L}$.
As pointed out in
Ref.~\cite{Belvedere:2024wzg}, such arguments
can lead to an underestimate in the sensitivity enhancement due to
various improvements in analyses that become possible because of the opening
up of new exclusive phase space regions. We shall nevertheless project the
expected sensitivity from the LHC Run II analysis at integrated
luminosity 
${\mathcal L}_0$ 
assuming an enhancement by a factor $\sqrt{{\mathcal
    L}/{\mathcal L}_0}$, bearing in mind that it is a rather conservative
\emph{under}estimate. We show the resulting estimate in Fig.~\ref{fig:val}b
for a 
narrow $Z^\prime$ by
scaling the ATLAS estimates of its own sensitivity in LHC Run
II~\cite{ATLAS:2019erb}. In the following sensitivity estimates for our model
line, we shall use the interpolation (\ref{interp}) in $z$, everywhere 
substituting expected sensitivities for $s$. 


CMS has performed similar searches to the ATLAS ones described
above~\cite{CMS:2021ctt}. 
The resulting
bounds on $Z^\prime$ bosons are similar to those of ATLAS and we therefore
expect similar exclusion limits\footnote{A CMS analysis~\cite{CMS:2023nzs} examines di-muon final states associated with $b-$jets,
which will also provide constraints on parameter space via the associated
$g b \rightarrow Z^\prime b$ production channel.}. 

\begin{figure}
  \begin{center}
    \unitlength=\textwidth
    \begin{picture}(0.99, 0.37)
      \put(-0.15,-0.1){\includegraphics[width=0.85 \textwidth]{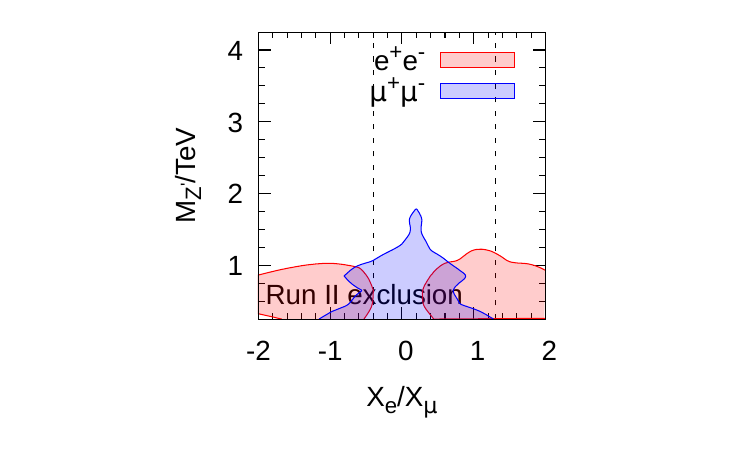}}
      \put(0.35,-0.1){\includegraphics[width=0.85 \textwidth]{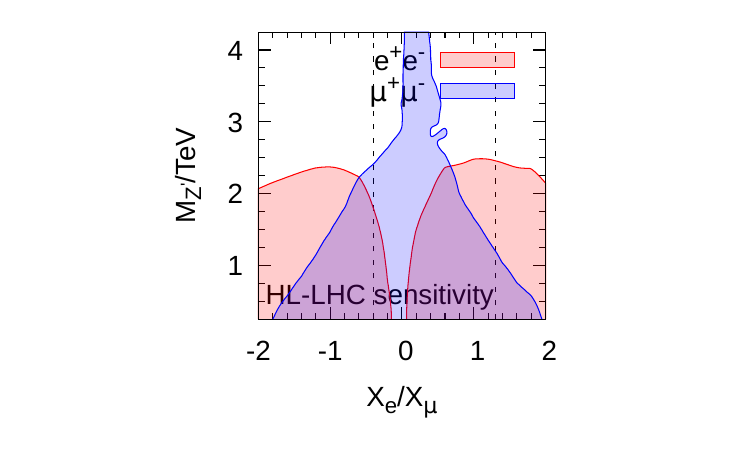}}
      \put(0.0,0.37){(a)}
      \put(0.5,0.37){(b)}
    \end{picture}
  \end{center}
  \caption{\label{fig:bounds} ATLAS di-lepton searches for $Z^\prime$ models
    that 
  fit \bsll\ data, for $X_\mu=10$.  $X_e/X_\mu$
  is the ratio of the charges of first family leptons to that of second
  family
  leptons under the additional spontaneously broken $U(1)_X$ gauge symmetry.
  At every point, $\theta_{sb}$
  and $g_{Z^\prime}$ were profiled by the fit to flavour and LEP-2 data as in Ref.~\cite{Allanach:2023uxz}.
  The 95$\%$ best-fit
  region of $X_e/X_\mu$ is the region between the two dashed lines. In (a),
  the filled regions are excluded at the 
  95$\%$ CL by the LHC Run II ATLAS di-lepton search for
  the channel   indicated in the legend. In (b), the filled regions denote our
  estimate of the HL-LHC sensitivity.
  }
  \end{figure}
In Fig.~\ref{fig:val}a, we display the bounds from the Run II ATLAS search. We
see that, depending upon the ratio\footnote{
For values of $|X_\mu|$ smaller than 10, $g_{Z^\prime}$ becomes larger at the
best-fit point, resulting in a larger $Z^\prime$ 
production cross-section, as in (\ref{prod}). 
The pertinent $Z^\prime$ branching ratios would change
according to  
Table~\ref{tab:pws} and (\ref{width}): one acquires smaller di-lepton
branching ratios for smaller $X_\mu$, $X_e$. 
The sensitivity would thus be modified by the two competing effects of
cross-section versus branching ratio.} $X_e/X_\mu$, the 95$\%$ lower CL limit
on $M_{Z^\prime}$ lies between 0.8 -- 1.8 TeV.
Fig.~\ref{fig:params}a shows that $M_{Z^\prime}$ has a maximum of between 10
and 35 TeV, depending upon $X_e/X_\mu$. We deduce that the direct 
search rules out less than a fifth of this 
parameter space.
In terms of $M_{Z^\prime}$, Fig.~\ref{fig:val}b shows that the HL-LHC is
expected to more than double the LHC reach,
whichever the value of $X_e/X_\mu \in [-2,2]$. 
Unless the
$Z^\prime$ of the model were to be discovered 
at the HL-LHC,
a more powerful collider (e.g.\ a 100 TeV 
$pp$ collider) would be 
needed to cover all of the remaining parameter space. 


\section*{Acknowledgements}
This work has been partially supported by STFC consolidated grants
ST/T000694/1 and ST/X000664/1. We thank the Cambridge Pheno Working Group for
helpful discussions and the Glasgow Particle Physics
Theory group for hospitality enjoyed while part of this work was carried out.  



  \bibliographystyle{elsarticle-num} 
  \bibliography{search.bib}


\end{document}